\documentclass{article}
\usepackage[T1]{fontenc}
\usepackage[utf8]{inputenc}
\usepackage{ismir} % Remove the 
\usepackage{amsmath,cite,url}
\usepackage{graphicx}
\usepackage{color}

\usepackage{csquotes}
\usepackage{booktabs}

\usepackage{adjustbox}
\usepackage{multirow}
\usepackage{array}
\usepackage{xcolor} 

\usepackage{needspace}

\usepackage{xargs}
\usepackage{xspace}

\newcommandx{\TODO}[1]{\colorbox{green}{TODO: #1}}
\newcommandx{\CHECK}[1]{\colorbox{yellow}{Check?: #1}}

\newcommand*{\eg}{e.g.}

\newcommand*{\maracatu}{\textit{Maracatu}}
\newcommand*{\pp}{\,\textrm{p.p.}\@\,}
\newcommand*{\vs}{\textit{vs.}}
\newcommand*{\tcnvone}{\texttt{TCNv1}}
\newcommand*{\tcnvtwo}{\texttt{TCNv2}}

\makeatletter
\renewcommand{\ttdefault}{pcr} % Ensures the same typewriter font
\renewcommand{\ttfamily}{\fontencoding{\encodingdefault}\fontfamily{\ttdefault}\selectfont\fontsize{\numexpr\f@size-1}{\numexpr\f@size+1}\selectfont}
\makeatother

\title{Towards Human-in-the-loop Onset Detection: A Transfer Learning Approach for \textit{Maracatu}}

\oneauthor
  {António Sá Pinto}
  {Faculdade de Engenharia da Universidade do Porto, Porto, Portugal\\
   INESC TEC, Porto, Portugal\\
   \texttt{asapinto@fe.up.pt}}

\def\authorname{A.~S.~Pinto}

\usepackage[bookmarks=false,pdfauthor={\authorname},pdfsubject={\pdfsubject},hidelinks]{hyperref}
\begin{document}

\maketitle

\begin{abstract}
  We explore transfer learning strategies for musical onset detection in the Afro-Brazilian \textit{Maracatu} tradition, which features complex rhythmic patterns that challenge conventional models. We adapt two Temporal Convolutional Network architectures: one pre-trained for onset detection (intra-task) and another for beat tracking (inter-task). Using only 5-second annotated snippets per instrument, we fine-tune these models through layer-wise retraining strategies for five traditional percussion instruments. Our results demonstrate significant improvements over baseline performance, with F1 scores reaching up to 0.998 in the intra-task setting and improvements of over 50 percentage points in best-case scenarios. The cross-task adaptation proves particularly effective for time-keeping instruments, where onsets naturally align with beat positions. The optimal fine-tuning configuration varies by instrument, highlighting the importance of instrument-specific adaptation strategies. This approach addresses the challenges of underrepresented musical traditions, offering an efficient human-in-the-loop methodology that minimizes annotation effort while maximizing performance. Our findings contribute to more inclusive music information retrieval tools applicable beyond Western musical contexts.
\end{abstract}

\section{Introduction}\label{sec:introduction}

%Normal font size: \the\fontdimen6\font
%Typewriter font size: {\ttfamily \the\fontdimen6\font}
%Normal font: \fontname\font
%Typewriter font: {\ttfamily \fontname\font}

Accurately identifying the precise moment when a musical note begins remains one of the fundamental challenges in audio signal processing. This task, known as musical onset detection, serves as a cornerstone for numerous Music Information Retrieval (MIR) applications.
Onset detection has historically been essential for rhythmic analysis, notably in beat tracking systems~\cite{Goto1994,Dixon2001a,Dannenberg2005}. While end-to-end learning models have recently bypassed this explicit step in some contexts, onset detection continues to be critical for diverse applications such as score following~\cite{Bock2012}, music segmentation~\cite{Pons2017}, and polyphonic music transcription~\cite{Vogl2017}.

The methodological evolution of onset detection mirrors broader trends in MIR research. Early approaches relied on signal processing techniques to identify significant changes in audio properties~\cite{Bello2005,Dixon2006}, followed by the introduction of feature-based machine learning methods~\cite{Marolt2002,Lacoste2006}. The field then shifted toward neural network architectures, beginning with Recurrent Neural Networks\,(RNNs)\cite{Eyben2010} and advancing to Convolutional Neural Networks\,(CNNs)\cite{Schluter2013}, which extract relevant features directly from raw audio or spectral representations.
Despite impressive advances in performance metrics (with top models achieving F1 scores approaching 90\% in recent evaluations\footnote{MIREX 2018, at \url{https://nema.lis.illinois.edu/nema_out/mirex2018/results/aod/}}), significant challenges persist in onset detection. In particular, accurately detecting soft onsets remains difficult even for advanced models~\cite{Tomczak2023}. Moreover, these data-driven approaches introduce additional challenges related to training data requirements and generalizability.

The effectiveness of supervised learning models hinges on the quality and diversity of training data~\cite{Peeters2021}. 
Current systems experience performance drops when analysing non-Western musical traditions or rare instruments, primarily due to insufficient representation in existing datasets. 
Addressing these gaps requires costly annotation efforts that demand both domain-specific and culturally-informed expertise~\cite{Srinivasamurthy2014a}, further complicating dataset curation. Furthermore, the annotation process itself reveals limitations: manual labelling of onsets is prone to human error and inconsistencies~\cite{Bolt2023}, with even isolated percussive signals proving difficult to label precisely~\cite{Davies2020}. 
These constraints restrict the practical deployment of state-of-the-art systems in diverse musical contexts, pointing to the need for more adaptable strategies.

Moving beyond the specific challenges of onset detection, MIR research has employed several adaptive strategies within rhythm analysis tasks. Informed methods leverage a priori knowledge about rhythmic content for tasks such as beat tracking~\cite{Fuentes2019a} and metre determination~\cite{Srinivasamurthy2017}, which, while effective in specific genres, lack generalizability. Transfer learning leverages knowledge across domains, with examples including adaptations of mainstream beat-tracking models to Greek folk music~\cite{Fiocchi2018} and facilitating adaptive rhythm microtiming generation~\cite{Burloiu2020}. Additionally, user-centric approaches like Active Learning and Few-Shot Learning optimize learning through strategic sample selection, enhancing adaptability in polyphonic drum transcription~\cite{Wang2020b,Wang2021} and enabling interactive refinement for onset detection~\cite{Valero2017} and beat tracking~\cite{Yamamoto2021}. This shift toward user involvement exemplifies the current human-centred landscape of MIR, recognizing users' essential role in data-driven systems~\cite{Schedl2014}. The integration of human expertise into computational frameworks provides a promising avenue when existing solutions prove insufficient.

Recent research has explored incorporating user-provided information to enhance beat tracking performance. Techniques such as high-level model parameterization~\cite{Pinto2019} and integrating user-annotated data snippets in a fine-tuning cycle~\cite{Pinto2021} have shown promise for improving state-of-the-art accuracy. These methods are particularly effective in addressing challenges in underrepresented musical contexts, where conventional MIR techniques underperform. Such approaches have proven instrumental in the creation of the \maracatu{} onset dataset~\cite{Davies2020}, metre determination in Latin-American music~\cite{Maia2022}, and beat tracking in highly challenging music signals~\cite{Pinto2023}.
The implementation of transfer learning for these tasks varies considerably: while some approaches retrain only final layers to leverage basic rhythmic representations~\cite{Choi2017a, Fiocchi2018}, others target input and output layers for instrument-specific adaptation~\cite{Davies2020}, and some retrain entire networks~\cite{Pinto2021}. Despite these varied strategies, no studies have empirically evaluated the impact of layer-wise retraining on model performance, leaving this critical question unexplored.

%It is well-known that as the depth of a neural network increases, the features it learns become progressively more abstract~\cite{Karpathy2015}. Through transfer learning, the initial layers, which capture basic representations of the musical rhythm, can be reused for related tasks. This has led to the practice of retraining only the final layers of the network~\cite{Choi2017a, Fiocchi2018}.
Building on this foundation, this paper explores a user-driven transfer learning approach for onset detection, focusing on the Afro-Brazilian tradition of \maracatu{}. We use the eponymous dataset~\cite{Davies2020}, which features complex rhythms and unique instrumental acoustic characteristics that cause leading models to struggle with achieving satisfactory performance.

Our methodology involves adapting a deep neural network for each instrument in the \enquote{terno}, the percussion ensemble central to \maracatu{}'s rhythm, based on a short annotated snippet per instrument. Through these instrument-specific adaptations, we demonstrate an effective and straightforward method to enhance state-of-the-art performance.
We investigate two distinct transfer learning scenarios: one with a model initially trained for onset detection, and another novel approach adapting a beat-tracking model to onset detection. This extends previous research~\cite{Davies2020,Fonseca2021} by exploring cross-task feature transferability and leveraging more complex models trained on larger datasets. Furthermore, we systematically evaluate layer-wise retraining strategies, examining the effectiveness of freezing different layer groups to identify optimal configurations for \maracatu{} onset detection.

\section{Methodology}
\label{sec:methodology}

Our approach addresses the limitations of existing models in non-mainstream signals by integrating user-provided short annotated snippets. We adapt the human-in-the-loop method proposed by Pinto et al. for beat tracking~\cite{Pinto2019,Pinto2021,Pinto2023} to the task of onset detection, leveraging state-of-the-art models through \textit{in-situ} fine-tuning. This user-centred methodology eliminates the need for extensive training from scratch, enabling end-users to swiftly obtain high-quality onset estimates that align with their judgments.

For onset detection in monotimbral signals, we adapt neural networks to each instrument's unique acoustic characteristics using just a single 5-second annotated snippet per instrument as the fine-tuning target. This approach demonstrates both minimal annotation effort and rapid adaptation cycles, yielding instrument-specific networks optimized for their corresponding acoustic properties while remaining computationally feasible for standard resources.
While our method is applicable to various DNN architectures, this study employs Temporal Convolutional Network (TCN)-based models for their efficient retraining capabilities. The TCN's performance in onset detection tasks is comparable to state-of-the-art models, as demonstrated in Section~\ref{sec:baselines}, making it suitable for our investigation.

We explore two transfer learning scenarios: an intra-task setting using a TCN onset detection model~\cite{Fonseca2021} and an inter-task setting that adapts a TCN beat tracking model~\cite{Bock2020} to onset detection. This inter-task approach can be framed as a domain adaptation problem, where a model trained for beat tracking is repurposed for onset detection. Given the inherent relationship between beats and onsets, this adaptation may benefit from the typically broader training data available for beat tracking models. To the best of our knowledge, this is the first study to explore domain adaptation from beat tracking to onset detection.

Furthermore, onset detection's unambiguous objective, when contrasted with the multifaceted nature of beat tracking, allows for clearer adaptation targets and, consequently, more straightforward interpretation of results. This motivated us to extend previous research by examining layer-wise retraining strategies. We systematically freeze different segments of the 15-layer TCN architectures, from the initial convolutional layers with small receptive fields to the deeper layers with larger dilation rates and wider receptive fields.
In total, our experimental cycle comprises 150 fine-tuning cycles (15 layer configurations\,$\times$\,5 instruments\,$\times$\,2 models). Through this comprehensive evaluation, we aim to investigate feature transferability between related rhythm analysis tasks and systematically assess the impact of different layer freezing configurations. 

In line with open science principles~\cite{Mcfee2019}, we provide a GitHub repository with our code and detailed results, including per-file evaluation metrics for all configurations and higher-resolution figures for detailed analysis\footnote{\url{https://github.com/asapsmc/HIILOnsetDetection}}. 
The remainder of this section outlines the \maracatu{} dataset composition, experimental settings, base models' description, and fine-tuning and evaluation details.

\subsection{Dataset}

\textit{Maracatu de baque solto}\footnote{Hereafter referred to as \textit{Maracatu}, this genre should be distinguished from \textit{Maracatu de baque virado} (or \enquote{Na\c{c}\~ao}). Both share African origins and certain musical similarities, but differ significantly in instrumentation, practice, and narrative~\cite{Santos2009}.}, also known as \textit{Maracatu} \enquote{rural}, is a vibrant carnival performance from Pernambuco, Northeast Brazil, combining music, poetry, and dance~\cite{BessonieSilva2021}. The rhythmic nucleus of \textit{Maracatu}, known as the \enquote{terno} ensemble, consists of five percussionists playing traditional handmade instruments: \textit{cuica}, \textit{gonge-lo}, \textit{tarol}, \textit{mineiro}, and \textit{tambor-hi}. The \maracatu{} dataset~\cite{Davies2020} captures these instruments using contact microphones for largely isolated per-instrument tracks, recorded during a fixed location performance and comprising 34 individual pieces totalling approximately 33 minutes\footnote{While the original dataset contains 34 files per instrument, we excluded \textit{Instrument\_34} files across all sub-datasets due to a corrupted \textit{Mineiro\_34} file.}.

\textit{Maracatu} features two main rhythmic patterns: \enquote{marcha} and \enquote{samba}, characterized by fast tempi of approximately $165$ and $180$ beats-per-minute (bpm), respectively. 
This rapid pace creates a complex timing profile across the ensemble. Time-keeping instruments (\textit{cuica} and \textit{gonge-lo}) maintain rhythmic stability despite their sporadic use, with a mean onset count of around $4,700$ (2.5 annotations per second). In contrast, the \enquote{voicing} instruments (\textit{tarol}, \textit{mineiro}, and \textit{tambor-hi}) play more expressive roles, resulting in a higher mean onset count of approximately $16,600$ (8.9 annotations per second).

\begin{figure}[h]
  \centering
  \hspace*{-0.017\textwidth} 
  \includegraphics[width=.45\textwidth]{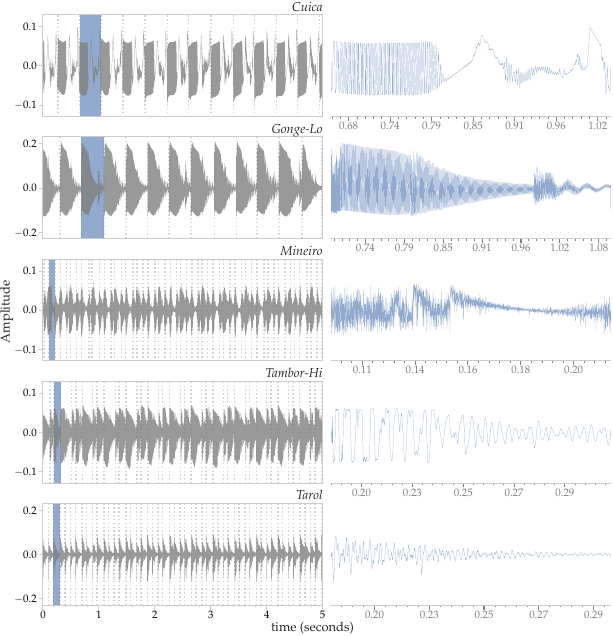}
  \caption{Onset-annotated waveforms for the \textit{Maracatu} instruments. Left: $5$-seconds fine-tuning snippet; Right: Zoomed in waveform, from the second onset to the sample before third onset (in \textit{blue}).}
  \label{fig:onsetdetection_maracatu}
\end{figure}

The intricacy of these rhythms and distinct waveform shapes, as illustrated in Figure~\ref{fig:onsetdetection_maracatu}, complicates onset detection and annotation. The \textit{mineiro} exemplifies this challenge with its unusual waveform characteristics, which led to its exclusion from microtiming analysis in the original dataset creation study due to annotation difficulties~\cite{Davies2020}. Combined with the under-representation of these instruments in available model training data, these factors create substantial obstacles for both human annotators and automated systems. The \textit{Maracatu} dataset thus provides an ideal test bed for our human-in-the-loop strategy, extending the approach previously employed in the dataset's creation.

\subsection{Base Models}
This study employs two pre-trained models, both derived from the TCN architecture proposed by Davies and B{\"{o}}ck~\cite{Davies2019}. For the intra-task setting, we use a modified version of the original TCN model with an additional 11th dilation rate level~\cite{Fonseca2021}, trained from scratch on the \textit{OnsetDB} dataset~\cite{Bock2012} for onset detection. In the inter-task scenario, we utilize an adaptation of the~\cite{Bock2020} multitask network, modified by masking its tempo and downbeat loss to function as a single-task (beat) network, trained on various beat-tracking datasets. Hereafter, we refer to these models as \tcnvone{} and \tcnvtwo{}, respectively.

\begin{figure}[h]
  \centering
  \includegraphics[width=.33\textwidth]{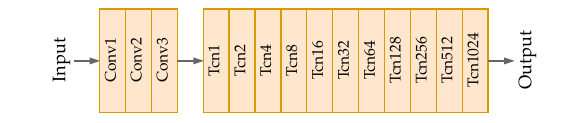}
  \caption{High-level architecture shared by the \tcnvone{} and \tcnvtwo{} models. Both follow the same layer sequence and depth, but differ in convolutional filter configuration, resulting in distinct receptive fields and overall model sizes.}
  \label{fig:TCN_architectures}
\end{figure}

As illustrated in Figure~\ref{fig:TCN_architectures}, both models share the same high-level architecture and signal conditioning stages, but their implementations differ significantly. 
\tcnvone{} consists of three convolutional layers with 16 filters and filter shapes of 3$\times$3, 3$\times$3, and 1$\times$8, with max pooling over three frequency bins after the first two layers. In contrast, \tcnvtwo{} employs three convolutional layers with 20 filters and filter shapes of 3$\times$3, 1$\times$10, and 3$\times$3, each followed by max pooling over three frequency bins. Both architectures use dropout after each convolutional stage. The ensuing TCN block operates non-casually and consists of 11 dilation levels, 16 filters, and a kernel size of 5. The \tcnvone{} model comprises 21,890 parameters, while the \tcnvtwo{} model has 116,302 parameters.
The original training procedures also differed slightly in optimization techniques: \tcnvone{} employed a standard \textit{Adam} optimizer, whereas \tcnvtwo{} used a \textit{Rectified Adam} plus \textit{Lookahead} approach.

\subsection{Fine-tuning}
For both intra-task and inter-task transfer learning settings, we adopt the fine-tuning strategy described in~\cite{Pinto2021}, using a 5-second annotated sample per instrument to demonstrate minimal annotation effort. Each base model is fine-tuned for 50 epochs with the learning rate reduced to one-quarter of the original value, maintaining the original optimizers for seamless training continuation. 
Early stopping and learning rate reduction mechanisms were not implemented as these parameters proved sufficient for convergence given the short training duration and small dataset size. Given our systematic layer-wise analysis comprising 150 fine-tuning cycles, we omitted data augmentation and additional hyperparameter optimization to maintain experimental tractability and support isolated analysis of how layer-wise freezing strategies relate to each instrument's acoustic characteristics.

We evaluate all possible fine-tuning configurations, denoted as \texttt{ft\textsubscript{A-B}}, where A and B indicate the starting and ending layers of the frozen section, respectively. The output layer is always updated and thus excluded from this notation. We explore configurations from \texttt{ft\textsubscript{Conv1$\ldots$3}} to \texttt{ft\textsubscript{Tcn1$\ldots$1024}}, including the fully trainable configuration \texttt{ft}. These are compared with the intra-task baseline \texttt{bsl} and the inter-task baseline \texttt{bsl\textsuperscript{*}}. 

\subsection{Evaluation}
The network output is an onset activation function with a 10-millisecond (ms) temporal resolution. We apply the standard \texttt{madmom} peak-picking algorithm to obtain onset estimates. Performance is evaluated using the F1 metric with the default $25$\,ms tolerance window~\cite{Bock2012}. 
We implement a holdout validation approach where, for each instrument, we extract a 5-second segment from the first file (\textit{Instrument\_01}) for fine-tuning and then exclude this entire file from the evaluation set to prevent data leakage. This ensures unbiased assessment of the instrument-adapted models by evaluating performance on the remaining 32 files per instrument.

\section{Experiments and Results}

\subsection{Preliminary Model Analysis}
\label{sec:baselines}
To contextualize our approach, we first compare the performance of our base TCN models with previous state-of-the-art methods on the \textit{OnsetDB} dataset~\cite{Bock2012}. 
Our base models, \tcnvone{} and \tcnvtwo{}, achieve F1 scores of 0.907 and 0.340, respectively. The lower performance of \tcnvtwo{} is expected, as it was originally trained for beat tracking rather than onset detection. The \texttt{madmomRNN} and \texttt{madmomCNN} models, pre-trained and provided as inference-ready models in the \texttt{madmom} package~\cite{Bock:2016_Madmom}, achieve F1 scores of 0.849 and 0.913, respectively. 
However, it is important to note that these evaluations were conducted without knowledge of the original training/test splits used for these pre-trained models, creating potential data leakage that may lead to an overestimation of their performance.
The 2nd generation onset CNN~\cite{Schluter2014} remains the established benchmark, with a reported F1 score of 0.903, verified through k-fold cross-validation. Unlike the \texttt{madmom} models, our TCN models were evaluated under the same validation conditions as the 2nd gen CNN, ensuring comparability. These results indicate that \tcnvone{} is competitive with the current state of the art in onset detection.

\subsection{\textit{Onset-to-Onset} Transfer Learning Results}

\begin{figure*}[ht!]
  \centering
  \includegraphics[width=\linewidth]{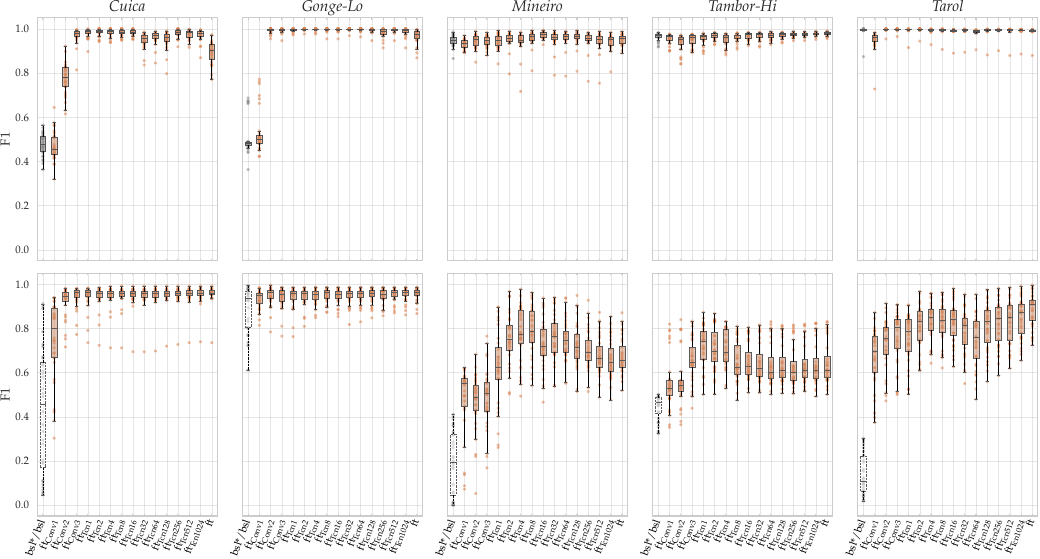}
  \caption{Distribution of F1 scores per layer-wise configuration under two transfer learning settings: \textit{Onset-to-Onset} (top), where fine-tuned models are compared against their baseline, and \textit{Beat-to-Onset} (bottom), where we assess cross-task versus within-task transfer learning, with comparable performance observed for time-keeping instruments.}
  \label{fig_boxplot_maracatu}
\end{figure*}

Figure\,\ref{fig_boxplot_maracatu} (top) presents the F1 scores obtained for each fine-tuning configuration in comparison to the baseline. The results can be grouped based on the rhythmic role of the instruments: time-keeping (\textit{cuica} and \textit{gonge-lo}) \vs{} voicing (\textit{tarol}, \textit{mineiro}, and \textit{tambor-hi}).

For time-keeping instruments, the baseline performance is moderate (F1 \ensuremath{\approx} 0.5), but fine-tuning yields significant improvements, with scores reaching the 0.8--1.0 range. In contrast, expressive instruments exhibit higher initial F1 scores (\ensuremath{\approx} 0.9--1.0), which limits the relative improvement. This disparity can be attributed to the conventional nature of \textit{tarol} and \textit{tambor-hi}, which are more aligned with the training data, whereas \textit{cuica} and \textit{gonge-lo} diverge more in terms of acoustic characteristics. An exception is \textit{mineiro}, which achieves a relatively high baseline score despite its distinct waveform characteristics. However, the reported lower precision of these ground-truth annotations~\cite{Davies2020} complicates direct performance comparisons.

Table~\ref{tab:best_models_full} presents high-performing configurations to demonstrate the achievable improvements across instruments. The \texttt{ft\textsubscript{Tcn16}} model achieves the highest accuracy for \textit{cuica} and \textit{mineiro} (0.985 and 0.972, respectively), while \texttt{ft\textsubscript{Tcn2}}, \texttt{ft\textsubscript{Tcn4}}, and \texttt{ft\textsubscript{Tcn16}} all achieve the highest F1 score for \textit{gonge-lo} (F1 = 0.998). For \textit{tambor-hi}, the best performance is obtained with both \texttt{ft\textsubscript{Tcn1024}} and \texttt{ft} (F1 = 0.978). For \textit{tarol}, the highest F1 score (0.997) is achieved with \texttt{ft\textsubscript{Conv3}}, though many configurations show comparable performance with marginal differences. These configurations consistently outperform the baseline, with the most notable gains observed in \textit{cuica} and \textit{gonge-lo}, where F1 improvements exceed 50 percentage points (p.p.).

In summary, all instruments benefit from adaptation, as most fine-tuned configurations—and in particular, the best for each instrument—consistently outperform the baseline. The improvement is especially pronounced for time-keeping instruments (\textit{cuica} and \textit{gonge-lo}), likely due to their lower baseline accuracy, which allows more room for improvement, and the relative ease of detecting sparser onsets compared to those that are closely clustered in time, even though onset density remains well above the network’s temporal resolution of 10\,ms. The optimal freeze configuration varies by instrument, with no clear global trend. However, some patterns emerge: for voicing instruments, full-network fine-tuning (\texttt{ft}) ranks among the top-performing configurations, whereas it degrades performance for time-keeping instruments.

\begin{table}[]
  \footnotesize
  \centering
  \caption{Representative configurations demonstrating improvements across transfer learning settings.}
  \label{tab:best_models_full}
  \begin{tabular}{@{}llrrlrr@{}}
    \toprule
                       & \multicolumn{3}{c}{\textit{Onset-to-Onset}}   & \multicolumn{3}{c}{\textit{Beat-to-Onset}}                                                                                          \\
    \cmidrule(lr){2-4} \cmidrule(l){5-7}
                       & \multicolumn{2}{c}{Adapted (best)} & \texttt{bsl}                        & \multicolumn{2}{c}{Adapted (best)} & \texttt{bsl\raisebox{0.5ex}{*}}                   \\\midrule
    \textit{Cuica}     & \texttt{ft\textsubscript{Tcn16}}   & 0.985                               & {0.477}                            & \texttt{ft}                     & 0.955 & {0.429} \\
    %\midrule
    \textit{Gonge-Lo}  & \texttt{ft\textsubscript{Tcn2/4/16}}    & 0.998                               & {0.508}                            & \texttt{ft}                     & 0.956 & {0.892} \\

    \textit{Mineiro}   & \texttt{ft\textsubscript{Tcn16}}   & 0.972                               & {0.946}                            & \texttt{ft\textsubscript{Tcn8}} & 0.790 & {0.193} \\

    \textit{Tambor-Hi} & \texttt{ft\textsubscript{ /Tcn1024}}                      & 0.978                               & {0.965}                            & \texttt{ft\textsubscript{Tcn1}} & 0.723 & {0.443} \\

    \textit{Tarol}     & \texttt{ft\textsubscript{Conv3}}   & 0.997                               & {0.993}                            & \texttt{ft}                     & 0.884 & {0.139} \\

    \bottomrule
  \end{tabular}
\end{table}

\subsection{\textit{Beat-to-Onset} Transfer Learning Results}
In this section, we focus on a domain adaptation, where a model pre-trained for beat tracking is adapted for onset detection. Unlike the previous setting, the goal here is not to compare fine-tuned models to their baseline, as this originates from a different task. We also refrain from an in-depth analysis of mean F1 scores across datasets, given their limited interpretative value. Instead, we assess whether models fine-tuned in this setting achieve results comparable to those in the onset-to-onset transfer learning scenario. 
Figure\,\ref{fig_boxplot_maracatu} (bottom) provides an overview of the results.

Time-keeping instruments, such as \textit{cuica} and \textit{gonge-lo}, achieve relatively high baseline (\texttt{bsl\textsuperscript{*}}) accuracies, likely due to the alignment between their onsets and beat locations. Adaptation improves accuracy across all instruments, confirming the feasibility of beat-to-onset transfer learning. However, while the fine-tuned models consistently outperform the beat-tracking baseline, direct comparisons to the onset-to-onset setting reveal performance disparities that vary by instrument. Specifically, for time-keeping instruments, performance remains nearly identical across both transfer learning scenarios, with differences of only 1.6\pp for \textit{cuica} and 3.7\pp for \textit{gonge-lo}. In contrast, voicing instruments exhibit progressively larger discrepancies, with F1-score differences of $11.3$\pp for \textit{tarol}, $27.5$\pp for \textit{mineiro}, and the largest gap of $32.2$\pp for \textit{tambor-hi}. 

Closer inspection of the layer-wise results reveals additional patterns. The accuracy generally increases as more layers are fine-tuned up to the 3rd or 4th dilation level, beyond which no further gains are observed. However, this trend does not hold for \textit{tarol}, where deeper fine-tuning leads to additional performance improvements. These observations highlight that, while fine-tuning is beneficial across all cases, the optimal retraining depth remains instrument-dependent.

Altogether, the results indicate that feature transferability from beat tracking to onset detection is more effective for time-keeping instruments than for voicing instruments. 
Specifically, \textit{gonge-lo} exhibits a clearly higher baseline F1 accuracy in the beat-to-onset setting compared to its onset-to-onset counterpart (0.892 vs. 0.508), while \textit{cuica} achieves a comparable performance (0.429 vs. 0.477), as reported in Table\,\ref{tab:best_models_full}.
This enhanced cross-task adaptability arises from the metrical function of time-keeping instruments: their onsets inherently coincide with beat positions, making them natural targets for the pre-trained model’s rhythmic representations.
Examining these results more closely, we verify that \textit{Maracatu}'s tempo range of 165–180 BPM corresponds to inter-beat intervals of 333–363\,ms. These durations approximately match the waveform spans of \textit{cuica} and \textit{gonge-lo}, but not those of the other instruments\footnote{According to an informal inspection of waveform spans---\textit{cuica}: 384-428 ms, \textit{gonge-lo}: 376-400 ms, \textit{tarol}: 77-107 ms, \textit{mineiro}: 90-180 ms, and \textit{tambor-hi}: 120-230 ms.}.
 
This temporal alignment—where the instruments' acoustic profile align with the genre's inter-beat intervals—explains the high baseline accuracies.
Additionally, the larger capacity of \tcnvtwo{} (116,302 parameters vs. 21,890 in \tcnvone{}) and its exposure to a broader training set may further contribute to this advantage. 
This suggests that model expressivity and pre-training diversity can compensate for task differences in certain transfer learning scenarios.

\subsection{Discussion}
Our investigation of two contrasting transfer learning scenarios reveals that adaptation outperforms baseline approaches across all instruments, with varying degrees of improvement. 

In the within-domain setting, adaptation yielded high accuracies with F1 scores from 0.972 (\textit{mineiro}) to 0.998 (\textit{gonge-lo}) and 0.997 (\textit{tarol}). Improvement was most pronounced for time-keeping instruments with lower baseline accuracies (\ensuremath{\approx}\,0.500), with \textit{cuica} showing a 52\pp gain.
For the cross-domain adaptation, while improvements over the beat-tracking baseline (\texttt{bsl\textsuperscript{*}}) were evident, comparison against the onset-tracking baseline (\texttt{bsl}) revealed instrument-dependent patterns. Voicing instruments' best F1 scores remained below the onset-tracking baseline by 11-24\pp, indicating limited benefits from domain adaptation. However, for time-keeping instruments whose onsets align with the pre-trained model's rhythmic priors, cross-task adaptation yielded improvements of 45-48 percentage points.

These findings provide key insights: i) Fine-tuning consistently enhances performance in both settings, making it valuable for achieving high accuracy in underrepresented music genres; ii) Models trained on beat-tracking can be effectively adapted for onset detection, leveraging model scale to compensate for task divergence and addressing limited data availability for non-mainstream instruments. However, effectiveness varies by instrument type: beat-to-onset adaptation benefits time-keeping instruments, while onset-to-onset adaptation consistently improves performance across all instruments. These improvements are naturally more substantial when baseline accuracy is lower, as observed in voicing instruments.

Our results also demonstrate that optimal fine-tuning configurations vary by instrument, necessitating tailored strategies for selecting which layer weights to update during fine-tuning. This challenges the assumption that only layers closest to the musical surface and the output layer would require recalibration to optimize a network for a specific instrument~\cite{Davies2020}.

Finally, several limitations warrant consideration. Our results represent a single experimental cycle, and despite prior research suggesting relative stability across runs~\cite{Pinto2021, Pinto2023}, the stochastic nature of the (re)training process--due to convolutional dropout--implies that results may vary. While unlikely to affect general trends, multiple cycles would be needed to investigate specific aspects such as receptive field size impact and its relation to optimal layer freeze selection or instrument waveform profiles. 
Note that, as previously discussed, corresponding layers across the two models differ in their temporal receptive fields despite having the same labels. For instance, while \texttt{Conv3} corresponds to approximately 50\,ms in both models, the layer \texttt{Tcn2} spans 170\,ms in \tcnvone{} vs. 410\,ms in \tcnvtwo{}. This discrepancy must be considered when interpreting results, limiting direct comparison between specific freeze configurations across scenarios.
The lower annotation precision of \textit{mineiro} further limits some result interpretation, potentially explaining its anomalous performance (\eg{} lowest fine-tuned and baseline accuracy on each setting).

Notably, our current results were achieved with minimal adjustment to the experimental pipeline to maintain fair comparison with baselines. This conservative approach suggests greater improvements might be possible through hyperparameter optimization—for example, cross-task adaptation may require more epochs to converge than within-task adaptation. While such optimization exceeded this study's scope, it represents a promising direction for extending the clear performance gains demonstrated here.

%Note that although both networks share the same layer names, they represent different temporal receptive fields (e.g., the \texttt{Conv3} layer corresponds to \CHECK{10 and 15} ms in \tcnvone{} and \tcnvtwo{}, respectively), which prevented direct comparison between specific freeze configurations across scenarios. 

\section{Conclusion}

This study investigated onset detection in \textit{Maracatu de baque solto} through two transfer learning strategies: onset-to-onset adaptation and beat-to-onset adaptation. Both approaches yielded notable improvements over baseline models, underlining the advantages of fine-tuning for enhancing accuracy.

We demonstrated that cross-task adaptation of models is viable for less-represented tasks such as onset detection when structural alignment exists between source and target domains. Transfer learning effectively addresses limited data availability and circumvents extensive manual annotation or costly training from scratch—a finding with important implications for music information retrieval, particularly when facing data scarcity challenges.

Future work should address this study's limitations while exploring in greater detail the factors influencing transfer learning effectiveness. Multiple-run experiments would confirm observed trends and investigate specific aspects, such as optimal freeze segment selection and its relation with network receptive field and instrument waveform profiles, alongside potential improvements through hyperparameter optimization.
Additional research directions include extending the analysis to other datasets and underrepresented instruments, and refining training protocols. 
Evaluating our adaptive approach using stricter tolerance windows would provide deeper insights into temporal precision, particularly for expressive instruments and microtiming analysis applications where fine-scale temporal variations are significant.
%Evaluating our adaptive approach using stricter tolerance windows beyond the standard \textpm25\,ms threshold would provide deeper insights into temporal precision, particularly for expressive instruments and microtiming analysis applications where fine-scale temporal variations are significant.

In summary, this study demonstrates the effectiveness of transfer learning in improving musical onset detection for diverse traditions beyond the Western canon. By adapting existing models, we can improve accuracy and robustness for underrepresented sounds. These methods and insights contribute to developing more inclusive tools for music analysis, with applications extending beyond the specific genres and tasks studied here to benefit the broader field of Music Information Retrieval.

\bibliography{references}

\end{document}